\documentclass[fleqn,twoside,twocolumn,nofootinbib,showkeys]{revtex4} % Specifies the document class %,unsortedaddress
\usepackage[nocpr]{ujp} % \usepackage[cyr]{ujp} for cyrillic, \usepackage[web]{ujp} for web
\begin{document}
\title[Excitonic Parameters of In$_{x}$Ga$_{1-x}$As--GaAs
Heterostructures]%колонтитул
{EXCITONIC PARAMETERS OF In\boldmath$_{x}$Ga$_{1-x}$As--GaAs
HETEROSTRUCTURES WITH QUANTUM WELLS\\ AT LOW TEMPERATURES}%
\author{N.M. Litovchenko}%1 автор
\affiliation{V.E. Lashkaryov Institute of Semiconductor Physics, Nat. Acad. of Sci. of Ukraine}%институт
\address{41, Prosp. Nauky, Kyiv 03028, Ukraine}%адрес
\email{strilchuk@isp.kiev.ua}%e-mail
\author{D.V. Korbutyak}%
\affiliation{V.E. Lashkaryov Institute of Semiconductor Physics, Nat. Acad. of Sci. of Ukraine}%
\address{41, Prosp. Nauky, Kyiv 03028, Ukraine}%
\email{strilchuk@isp.kiev.ua}
\author{O.M.~Strilchuk}
\affiliation{V.E. Lashkaryov Institute of Semiconductor Physics, Nat. Acad. of Sci. of Ukraine}%институт
\address{41, Prosp. Nauky, Kyiv 03028, Ukraine}%адрес
\email{strilchuk@isp.kiev.ua}

\udk{538.911, 538.958} \pacs{68.35.Ja, 78.55.Cr,\\[-3pt] 78.67.De} \razd{\secviii}

\autorcol{N.M.\hspace*{0.7mm}Litovchenko,
D.V.\hspace*{0.7mm}Korbutyak, O.M.\hspace*{0.7mm}Strilchuk}

\setcounter{page}{260}%

%\maketitle

%\makeatletter
%\renewcommand{\thesection}{\arabic{section}}
%\renewcommand{\p@subsection}{}
%\renewcommand{\thesubsection}{\arabic{section}.\arabic{subsection}}
%\renewcommand{\p@subsubsection}{}
%\renewcommand{\thesubsubsection}
%{\arabic{section}.\arabic{subsection}.\arabic{subsubsection}}
%\makeatother

%\input{tcilatex}

\begin{abstract}
Characteristics of GaAs/In$_{x}$Ga$_{1-x}$As/GaAs heterostructures
with a single quantum well, which were obtained at various growth parameters,
are evaluated according to the results of measurements of
low-temperature photoluminescence (PL) spectra and their
corresponding theoretical analysis. The experimentally obtained
temperature dependences of the energy position of the PL band maximum,
$h\nu _{\max}$, band half-width, $W_{0}$, and intensity, $I$,
are examined. The values of energy of local phonons,
$E_{\mathrm{ph}}$, exciton binding energy, $E_{\mathrm{ex}}$,
and the Huang--Rhys factor, $N$, are determined. A comparison
between the values obtained for those quantities and the growth
parameters of considered specimens allowed us to assert that the
highest-quality specimens are those that are characterized by low
$N$ values and one-mode phonon spectra.
\end{abstract}

\maketitle

\section{Introduction}

In$_{x}$Ga$_{1-x}$As--GaAs heterostructures are widely used in
modern optoelectronics as structures that are capable of being
adapted for the convenient reception, transmission, and
transformation of radiation in various spectral ranges. Special
attention is attracted to heterostructures with quantized layers
owing to their enhanced sensitivity and a possibility to
additionally vary the optical spectrum \cite{1,2,3,4,5,6,7}. The
energy of an emitted quantum in such heterostructures is governed by
the distance between the size-quantization levels of electrons and
holes, $E_{e1\mbox{-}hh1}$, which, in turn, depend on the quantum
well (QW) width $d$ and the composition of a substitutional solid
solution ($x$ is the indium content). For instance, for the typical
values $x=0.2$ and $d=$ $=80\;\mathrm{\mathring{A}}$, the changes of
$x$ by 1\% and the well width by the width of a monolayer
(approximately 3 \AA) give rise to the variations by 9 and 4 meV,
respectively, in the transition energy \cite{4}. Such a high
sensitivity to the parameters imposes strict requirements on both
the width and the element composi- \mbox{tion of QW.}\looseness=1

%Tabl.1
\begin{table*}[!]
\vspace*{4mm} \noindent\caption{  }\vskip3mm\tabcolsep9.6pt

\noindent{\footnotesize\begin{tabular}{|c|c|c|c|c|c|c|c|c|}
  \hline
  \multicolumn{1}{|c}{\parbox{1.3cm}{\vspace*{6mm}Specimen}} &
  \multicolumn{1}{|c}{\parbox{0.6cm}{\vspace*{6mm}$x$ In} }
  &\multicolumn{1}{|c}{\parbox{0.7cm}{\vspace*{6mm}$d$, {\AA}}}
  &\multicolumn{1}{|c}{\parbox{1.0cm}{\vspace*{6mm}cap, {\AA}}}
  &\multicolumn{3}{|c}{\rule{0pt}{5mm}$T=5$}
  &\multicolumn{1}{|c}{\parbox{1.3cm}{\vspace*{6mm}$E_{a1}$, meV}}&
\multicolumn{1}{|c|}{\parbox{1.3cm}{\vspace*{6mm}$E_{a2}$, meV}} \\%[1mm]
\cline{5-7} &&&&\multicolumn{1}{|c}{\rule{0pt}{5mm}$h\nu _{\max}$,
eV}
  &\multicolumn{1}{|c}{$W$, meV}
  &\multicolumn{1}{|c|}{$I_{\max}$,  rel. units}&& \\[2mm]
\hline
\rule{0pt}{5mm}No. 1& 0.16& 84& 220& 1.3568& 7.4& 1093& 4.5&52 \\
No. 2& 0.21& 88& 230& 1.33& 9.9& 582& 2.8&55 \\
No. 3-1& 0.20& 92& 600& 1.355& 7.3& 4876& 5&70 \\
No. 3-2& 0.20& 92& 600& 1.3721& 10.9& 949& 2.5&46 \\
No. 4-1& 0.35& 73& 600& 1.253& 15.2& 302& 2.5&85 \\
No. 4-2& 0.35& 73& 600& 1.396& 13.2& 988& 2.5&60 \\[2mm]
\hline
\end{tabular}
\vskip2mm\parbox{16.7cm}{\raggedright{N o t a t i o n: $x$ is the
relative content of indium, $d$ is the quantum well width, cap is
the thickness of the protective GaAs la- yer,~ $h\nu_{\max}$~ is the
maximum position in the PL spectrum, $W$ is the line half-width, $I$
is  the PL intensity, $E_{a1}$ and $E_{a2}$~ are the activation
energies. }}}\vspace*{-1mm}
\end{table*}

Moreover, a shortcoming of those heterostructures consists in a considerable
mismatch between the lattice constants in the epitaxial layer and the
substrate, which results in the emergence of substantial mechanical stresses
and the generation of numerous dislocations. Buffer layers or the formation
of quaternary alloys with the phosphorus additive usually reduce the
influence of those undesirable factors, but not completely. Therefore, there
arises the requirement in a non-destructive quantitative control over the
deformation and defect factors.

In this report, we pay attention to a possibility of using the interaction
between phonons and excited electrons (the Huang--Rhys factor $N$) for the
characterization of the imperfection degree. This parameter is determined on the
basis of the temperature behavior experimentally found for the exciton
photoluminescence band half-widths and subjected to the corresponding
theoretical analysis.

\section{Experimental Specimens and Technique}

Low-temperature (5--200~$ \mathrm{K}$) photoluminescence (PL)
researches are carried out with the use of
he\-te\-ro\-struc\-tu\-res with a single quantum well,
GaAs/In$_{x}$Ga$_{1-x}$As/GaAs. The specimens were grown up
following the MOCVD technology. They were characterized by various
contents of indium, various widths of In$_{x}$Ga$_{1-x}$As quantum
wells, and various thicknesses of protective GaAs layers (Table~1).
Luminescence was excited by a He--Ne laser (a quantum energy of
1.96~eV, and the radiation intensity $L=(3\times 10^{17}\div
10^{19}) $~quantum/(cm$^{2}$$\cdot$s)). To analyze the PL spectra,
we used an MDR-23 monochromator with the spectral resolution not
worse than 0.2 meV. The signal was registered with the use of a
cooled FEP-62 photoelectronic \mbox{multiplier.}

\section{Experimental Part}

In the photoluminescence spectra of researched specimens, we
observed the intensive bands, which correspond to the exciton
recombination \textit{e1-hh1} in the quantum well in the interval
$T=\left( 5\div 40\right) ~ \mathrm{K}$ and to the recombination of
free charge carriers in the quantum well in the interval $T=\left(
50\div 200\right)  ~\mathrm{K}$. This fact is confirmed by the
dependences of the radiation intensity on the excitation one,
$I(L)$. The main growth parameters of studied specimens and the
corresponding photoluminescence bands are quoted \mbox{in Table
1.}\looseness=1

In Fig. 1, \textit{a}, the normalized PL spectra of the specimens
under investigation obtained at $T=5 ~\mathrm{K}$ are depicted.
Attention should be paid to some spectral features. First of all, it
is the spread in the energy positions of PL bands, which stems from
different values of quantum well widths $d$ and indium contents $x$
in the In$_{x}$Ga$_{1-x}$As quantum well. However, even provided
that the corresponding values of $d$ and $x$ are identical, the
energy positions of PL bands can differ substantially. For instance,
for specimens 4-1 and 4-2 with the identical $x=0.35$ and
$d=73$~{\AA}, the energy positions of PL maxima are different (see
Table~1); namely, $h\nu _{\max}=1.396~\mathrm{eV}$ for specimen 4-2
and 1.253~$ \mathrm{eV}$ for specimen 4-1. In our opinion, the
origin of such a discrepancy consists in fluctuations of the In
concentration in the QWs of specimens 4-1 and 4-2. Really, if the
energy difference between the radiation maxima, $\Delta h\nu
_{\max}=0.143~ \mathrm{eV}$, was caused by the difference between
the quantum well widths, the latter would be amount to about $107~
\mathrm{\mathring{A}}$ (this value can be obtained with regard for
the fact mentioned above that a change of the QW width by about
3~{\AA} gives rise to a variation of about 4~meV in the transition
energy), but such a value is unreal. This reasoning agrees with the
results of work \cite{4}, where it was noticed that a variation in
the spectral positions of exciton peaks for the In$_{x}$Ga$_{1-x}$As
quantum well is mainly associated with the variations in the In
 \mbox{concentration $x$.}\looseness=1

Different half-widths $W$ (Table 1) and shapes of examined PL
spectra exhibited in Fig.\,\,1,\,\,\textit{a }compose another
feature of those spectra. The PL band half-width depends on both the
degree of exciton localization in the QW and the character of
exciton scattering by phonons, defects, inhomogeneities at
heterointerfaces, and so forth. The radiation emission spectra of
some specimens demonstrate a characteristic tail of the PL band in
the low-energy interval, which may be caused by the participation of
phonons in the radiative recombination of excitons in the QW. As an
example, we decomposed the PL band of specimen 3-2 into two
components: the zero-phonon one and the phonon replica
(Fig.\,\,1,\,\,\textit{b}), with the use of the procedure proposed
in work \cite{8}. In so doing, we used an approximation that phonons
of only one type participate in PL. This approach enabled us to
obtain the value $E_{\rm ph}=9.8~\mathrm{ meV}$ for the energy of
interacting local phonons and $N=0.3$ for the Huang--Rhys factor,
which characterizes the strength of exciton--phonon
\mbox{interaction.}\looseness=1

%Fig. 1
\begin{figure}% figure* for wide figure, [h] [!] to change the placement
\vskip1mm
\includegraphics[width=7cm]{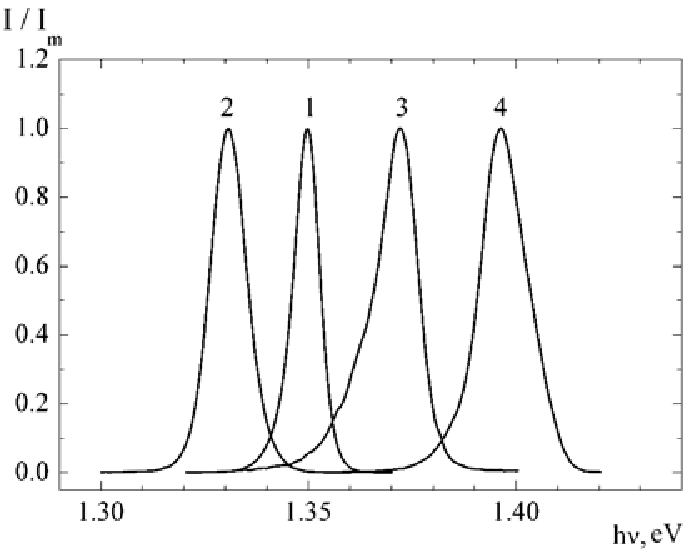}\\
{\it a}\\ [3mm]
\includegraphics[width=7cm]{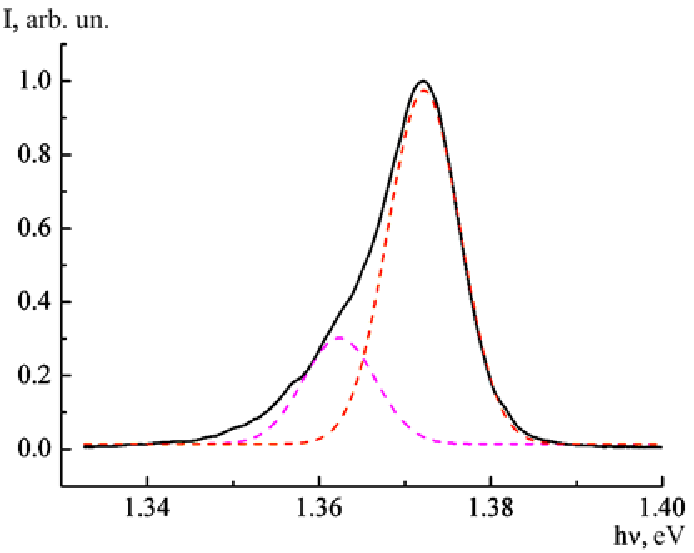}\\
{\it b} \vskip-3mm\caption{PL spectra of the studied
GaAs/In$_{x}$Ga$_{1-x}$As/GAs quantum heterostructures (\textit{a}).
De\-com\-po\-si\-tion of PL band 3 in panel \textit{a} in two
components (\textit{b}): the zero-phonon one  and the phonon
replica. $E_{\max}=1.372$~eV, $E_{\rm ph}=9.8$~meV, the Huang--Rhys
factor $N=0.3$  }\vskip3mm
\end{figure}

As to the half-width of PL bands (Table~1), it changes from 7.4~meV
for specimen 3-1 to 15.2~meV for specimen 4-1. In our case, the Bohr
radius of an exciton in the QW is comparable with the QW width.
Therefore, with a high probability, the exciton is localized at
inhomogeneities of heterointerfaces. The smaller width of the PL
band corresponds to a larger localization degree, as it takes place
for PL by bound excitons in a bulk semiconductor. The temperature
dependences of the PL intensity typical of the specimens under
consideration are shown in Fig.~2. At low temperatures ($T=5\div 40~
\mathrm{K}$), the PL intensity varies weakly. As the temperature
grows, the PL intensity decreases for specimens 1 and 2 and, for
specimens 3 and 4, first slightly increases and then falls down,
which is connected with the temperature-induced ejection of charge
carriers from the quantum well into the barrier.

In Fig. 3, the temperature dependences of the radiation maximum
position typical of examined specimens are depicted. The dashed
curve demonstrates the results of calculation obtained in the
framework of the Varshni model. For all specimens, the temperature
dependences of the PL maximum position have an S-like form. At low
temperatures, a considerable deviation of the calculated values from
experimental ones is observed: first, the position of the PL maximum
shifts toward low energies (red shift); then, up to a certain
temperature, the maximum shifts backward toward high energies. At
$T>(60\div 89)~\mathrm{K}$, the maximum position shifts toward low
energies in accordance with the \mbox{Varshni model.}\looseness=1

In Fig. 4, the typical dependences of the PL band half-width on the
temperature are shown. For specimens 1 and 2, a monotonous increase
of the half-width with the temperature (the dependence of type~I) is
observed. Specimens 4 are characterized by a sharp initial (to a
temperature of 40--80~$ \mathrm{K}$) growth of the half-width, then
by an insignificant reduction of this parameter followed by its
subsequent growth, as the temperature grows further (the dependence
of type~II). For specimens 3, the dependences of both types are
observed.

\section{Discussion of Experimental Results}

The temperature dependences of the PL intensity obtained for specimens 2
and 4-2 (Fig. 2) were analyzed with the use of the Arrhenius formula
\[
 I( T) =C/[ 1+a_{1}\exp(
{-E_{a1}/kT}) + a_{2}\exp ( {-E_{a2}/kT}) ].
\]

\noindent This enabled us to determine two temperature intervals
with different slopes: low- and high-temperature ones with the
activation energies $E_{a1}$ and $E_{a2}$, respectively (Table 1).
In the low-temperature interval, $E_{a1}=(2.5\div 5)~\mathrm{ meV}$.
Such a small value of $E_{a1}$ testifies that this quantity
corresponds to the delocalization energy of excitons bound at
inhomogeneities of heterointerfaces in the QW at low temperatures.
It is significant that the higher the delocalization energy (and,
accordingly, the deeper is the potential well, which is associated
with the corrugation of heterointerfaces), the narrower is the PL
band, which corresponds to a more localized state of excitons (see,
e.g., the PL bands for specimens 1 and 3-1). At temperatures $T>40
$~K, the activation energy for the temperature-induced quenching of
the PL band owing to the \textit{e1-hh1 }transitions equals
$E_{a2}=(50\div 85)~\mathrm{ meV}$, and, as was indicated above,
this is connected with the temperature-induced ejection of charge
carriers \mbox{into the barrier.}%\looseness=1

The temperature dependences of the PL maximum depicted in Fig. 3 and
their comparison with the results of calculations following the
Varshni formula allow the binding energy of excitons in the QW to be
evaluated. As one can see from Fig. 3, the energy of a quantum
emitted at low temperatures ($T=(5\div 40)~ \mathrm{K}$) is lower
than the energy of interband transitions calculated by the Varshni
formula. The corresponding difference $E_{\rm ex}\approx 10~\mathrm{
meV}$ is just the binding energy of excitons in the QW to within the
accuracy of the energy of exciton localization at
\mbox{heterointerfaces.}%\looseness=1

The theoretical analysis of the temperature dependences obtained for the
scattering parameter (the PL band half-width $W$) is based on the fact
that this quantity comprises the probability of the momentum scattering as a
result of several independent processes (by impurities, phonons, and
others),
%1
\[
W\sim \frac{\hbar }{\tau _{t}},
\]\vspace*{-7mm}
\begin{equation}
\frac{1}{\tau _{t}}\sim W_{t}\left( T\right)
+W_{\mathrm{ph.opt}}\left( T\right) +W_{\mathrm{ph.local}}\left(
T\right) +\mbox{...},  \label{eq1}
\end{equation}%

\noindent where $\tau _{t}$ is the lifetime of nonequilibrium charge
carriers. The temperature dependences of the scattering probability are
different for different mechanisms. Hence, the temperature
dependence can be used to distinguish between their contributions.
In particular, the Coulomb scattering by local centers depends on $T$,
which is the most pronounced at low enough temperatures. At the same
time, the role of the phonon mechanism grows with $T$
\cite{9,10}, according \mbox{to the law}
%2
\[
W=\Sigma  W_{\mathrm{oi}} \left(\! {\mathrm{cth}\frac{\hbar \omega
_{\mathrm{ph}} }{2kT}}\! \right)^{\!1/2}\!,
\]\vspace*{-5mm}
\begin{equation}
\label{eq2} W_{\mathrm{oi}} =2( {2\ln 2} )^{1/2} N_{\mathrm{phi}}
^{-1/2} \hbar \omega _{\mathrm{phi}} \sim N_{\mathrm{phi}}^{-1/2},
\end{equation}

%Fig. 2
\begin{figure}% figure* for wide figure, [h] [!] to change the placement
\vskip1mm\includegraphics[width=7cm]{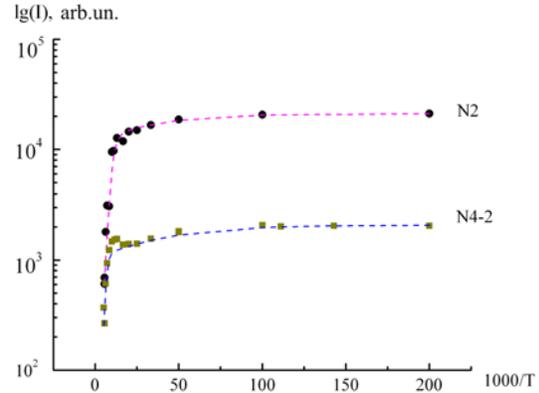}
\vskip-3mm\caption{Temperature dependences of the PL intensity:
symbols demonstrate experimental results, dashed curves correspond
to the approximation by the Arrhenius formula }\vskip3mm
\end{figure}

%Fig. 3
\begin{figure}%
\includegraphics[width=7cm]{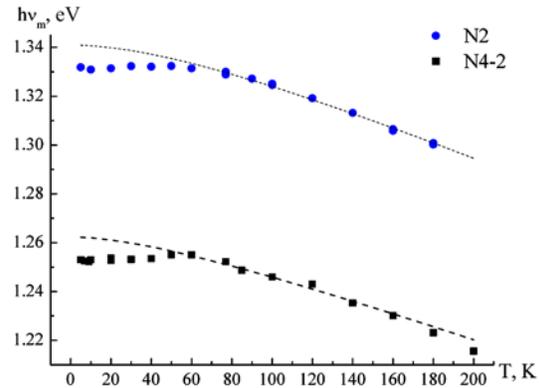}
\vskip-3mm\caption{Temperature dependences of the photoluminescence
maximum position, $hv_{m}$: symbols demonstrate the experimental
results, dashed curves show the results of calculations by the
Varshni formula  }\vskip3mm
\end{figure}

%Fig. 4
\begin{figure}% figure* for wide figure, [h] [!] to change the placement
\includegraphics[width=7cm]{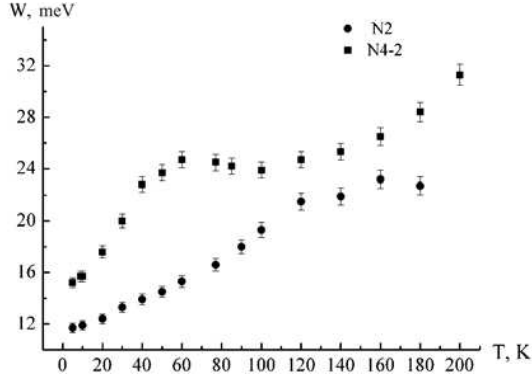}
\vskip-3mm\caption{Dependences of the photoluminescence band
half-width $W$ on the temperature  }
\end{figure}

\noindent
 where $E_{\mathrm{ph}}=h\omega _{\mathrm{ph}}$ is the
energy of a phonon localized at a radiative-recombination center, and
$N_{\mathrm{ph}}$ is the phonon emission probability at
the recombination (the Huang--Rhys factor). In the case $N<1$, the
latter is given by the following relation:
%3
\[
N\sim \frac{5}{8}\left( {E_{\mathrm{ex}}/E_{\mathrm{ph}}}\right)
 \left( {\varepsilon _{0}/\varepsilon _{\infty }-1}\right) \sim
\]\vspace*{-5mm}
\begin{equation}
\sim e^{2}/E_{\mathrm{ph}} \frac{1}{3a_{\rm B}}\left(\!
{\frac{1}{\varepsilon _{\infty }}-\frac{1}{\varepsilon
_{0}}}\!\right)\!, \label{eq3}
\end{equation}%
where $E_{\mathrm{ex}}$ is the exciton binding energy, $a_{\rm B}$
the Bohr radius of an exciton, and $\varepsilon _{0}$ and
$\varepsilon _{\infty }$ are the static and high-frequency
dielectric permittivities, respectively. From this formula and
knowing the phonon energy $E_{\mathrm{ph}}$, it is easy to find the
binding energy of an exciton~\cite{11},
%4
\begin{equation}
E_{\mathrm{ex}}\approx N E_{\rm ph}\, \frac{3}{2} \frac{\varepsilon
_{\infty }}{\varepsilon _{0}-\varepsilon _{\infty }}.  \label{eq4}
\end{equation}%

%Tabl.2
\begin{table}[b]
\vspace*{-1mm} \noindent\caption{  }\vskip3mm\tabcolsep8.0pt

\noindent{\footnotesize\begin{tabular}{|c|c|c|c|c|}
  \hline
  \multicolumn{1}{|c}{Specimen} &
  \multicolumn{1}{|c}{$E_{\rm ph}$, meV }&\multicolumn{1}{|c}{$N$}
  &\multicolumn{1}{|c}{$E_{\rm ex}$, meV}&
\multicolumn{1}{|c|}{\rule{0pt}{5mm}$a_{\rm B}$, {\AA}} \\[2mm]
  \hline \rule{0pt}{5mm}No. 1&8&0.15&9.1&122\\
  No. 2&6.5&0.42&20.8&54\\
  No. 3-1&11&0.08&6.7&168\\
  No. 3-2&8&0.27&16.5&8\\
  &12&0.149&13.6&82\\
  No. 4-1&3.2&3.1&78.7&14\\
  &18&0.13&19.0&58.8\\
  No. 4-2&3.5&3.4&95.4&12\\
  &19&0.22&33.5&33\\[2mm]
   \hline
\end{tabular}
  }
\end{table}

\noindent Hence, the temperature dependence $W(T)$ allows a number
of parameters that characterize the heterostructure state -- such as
$W_{0}$, $E_{\mathrm{ph}}$, and $N$ \cite{12} -- to be obtained, as
well as the Stokes shift,
\begin{equation*}
\Delta \omega_{\mathrm{st}}=2N_{\mathrm{ph}} \hbar \omega
_{\mathrm{ph}}.
\end{equation*}%
It also enables one to evaluate the position of the phonon-free line by
the formula $\hbar \omega _{0}=\hbar \omega _{\max }+n\hbar \omega
_{\mathrm{ph}}$, where $n$ is the number of phonon replica.

Special attention should be paid to the quantity $W$. Its magnitude is
reciprocal to the charge carrier mobility and is predicted to be much less
for a perfect quantum well than that for the corresponding bulk material.
However, mechanical stresses and defects can compensate this useful effect.
The proposed analysis allows the contributions of different mechanisms to be
estimated separately.

The features in the temperature dependences of the PL intensity,
maximum position, and half-width obtained in this work can be
explained by the presence of localized (defect) states in the
studied specimens \cite{6,7}, which are induced by fluctuations of
QW dimensions, and/or by a variation of the QW composition. At low
temperatures, photo-induced charge carriers (excitons) are captured
by the localized potential. As the temperature is elevated to a
value that corresponds to the localization energy maximum, a shift
of the PL maximum position toward lower energies (the red shift) is
observed, because excitons obtain a sufficient thermal energy to
overcome the potential barrier and become relatively free. Some of
those excitons relax into lower states, which capture them, and
recombine there. In this temperature interval, we observe a drastic
increase in the half-width $W$ of the PL band, in accordance with
the growth in the population of states owing to the capture of
released charge carriers onto them. As the temperature grows
further, the PL maximum shifts into the range of high energies, and
the band half-width becomes somewhat narrower due to the thermally
equilibrium distribution of excitons. This occurs until the
temperature corresponding to the complete delocalization of charge
carriers is attained. At higher temperatures, the \textit{e1-hh1}
transitions dominate in the PL spectrum, and, according to the
Varshni formula, the maximum position changes with the temperature
as the energy \mbox{gap width.}\looseness=1

With the use of Eqs. (\ref{eq2})--(\ref{eq4}) and the experimentally
obtained temperature dependences for the PL band half-width, we
determined the parameters $E_{\mathrm{ph}}$, $N$, $E_{\mathrm{ex}}$,
and the Bohr exciton radius $a_{\rm B}$ (see Table 2). Let us
consider this dependence of type I (it is inherent to specimens 1,
2, and 3-1). It has a monotonous character and can be described well
by Eq. (\ref{eq2}), in which the scattering processes with phonons
of energies 6 to 11 meV are taken into consideration
(Fig.~5,~\textit{a}).

For structures 4, the character of the band half-width dependence on the
temperature is of the other type (type II). For those specimens, we
determined two values for the energy of local phonons (Fig. 5):

(i) in the interval from 5 to 20~$ \mathrm{K}$, where a drastic
increase of the band half-width is observed, the energy of phonons
is 3.5--4 meV;

 (ii) in the interval from 30 to 200~$ \mathrm{K}$, an insignificant
narrowing of the PL band is observed, which is followed by the
increase of its half-width with the temperature; here, the energy of
phonons equals \mbox{18--19 meV.}

%Fig. 5
\begin{figure}% figure* for wide figure, [h] [!] to change the placement
\includegraphics[width=7cm]{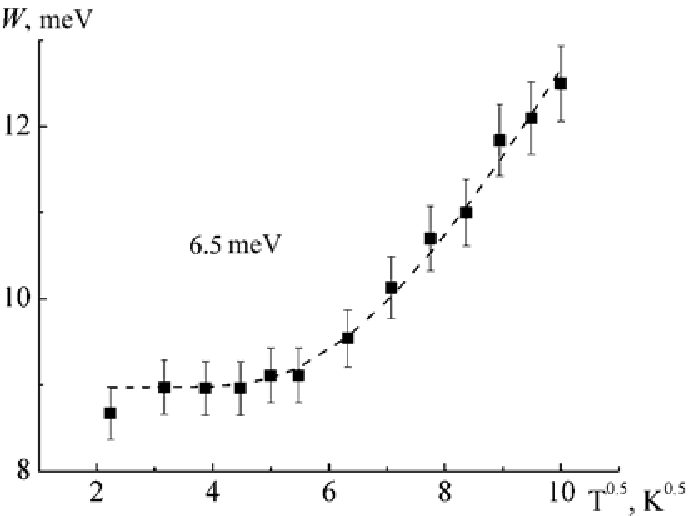}\\
{\it a}\\ [2mm]
\includegraphics[width=7.9cm]{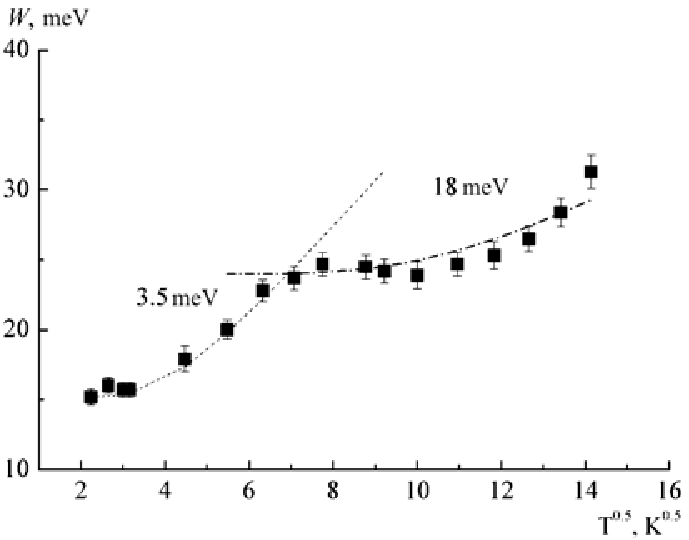}\\
{\it b}\\ [2mm]
\includegraphics[width=6.5cm]{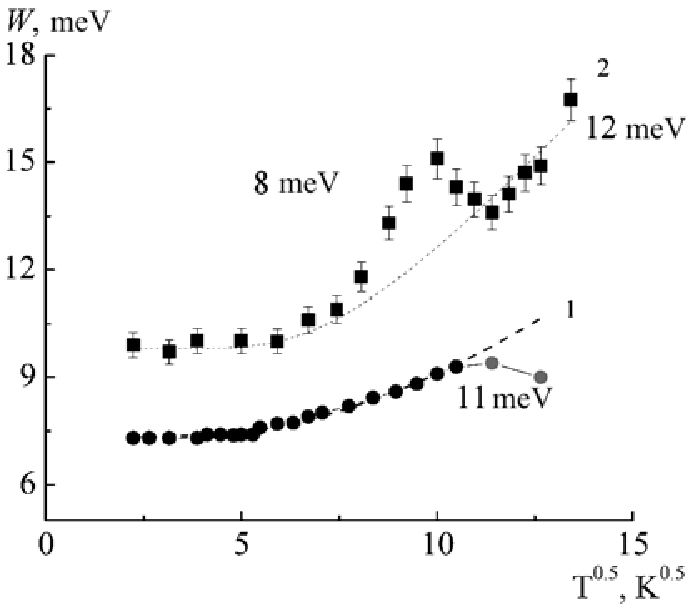}\\
{\it c} \vskip-3mm\caption{Temperature dependences of the PL band
half-width, $W(T^{0.5})$ for (\textit{a}) specimens 1 and 2,
(\textit{b}) specimen 4, and ({\it c}) specimens 3-1 and 3-2.
Symbols demonstrate experimental results, and dashed curves
correspond to their approximation by formula (\ref{eq2})  }
\end{figure}

Structures 3 revealed the dependences of both types. Specimen 3-1
demonstrated the dependence of monotonous type I, and the
corresponding energy of phonons was 11~meV (Fig. 5,\textit{c}, curve
\textit{1}). Specimen 3-2 was characterized by the dependence of
type II: the energy of phonons was 8~meV in the interval 5--60~$
\mathrm{K}$ and 12~meV in the interval 60--200~$ \mathrm{K}$
(Fig.~5,~\textit{c}, curve~\textit{2}).

The magnitudes of exciton binding energy obtained with the help of
Eq.\,\,(\ref{eq4}) for various specimens (\mbox{$E_{\rm
ex}>6.7~\mathrm{ meV}$}) considerably exceed the corresponding
energy for bulk excitons in In$_{x}$Ga$_{1-x}$As ($E_{\rm ex}\approx
3~\mathrm{ meV}$) \cite{1}, which testifies to the quantization of
excitons in the In$_{x}$Ga$_{1-x}$As--GaAs \mbox{quantum
well.}\looseness=1

Let us compare the values of $E_{\rm ph}$ obtained from the
temperature dependences of the line width with the theoretical
relations (\ref{eq1})--(\ref{eq4}) (see Table 2). For every
specimen, it turned out several times less than the characteristic
values for bulk or surface (confinement) phonons. An evident reason
is the fact that the studied specimens with heterojunctions had
rather a large number of defects, probably localized at interfaces.
It is known that one of the mechanisms of defect emergence in InGaAs
structures consists in the segregation of clusters of the In phase
if the optimum epitaxy and temperature regimes were not followed at
the stage of heterostructure formation \cite{3}. Hence, the typical
defects have to include precipitates of the redundant element,
\mbox{i.e. indium.}\looseness=1

Now let us estimate the energy of local phonons that correspond to
vibrations in vicinities of defects at the interfaces between InGaAs
and In precipitates.\,\,It is known that the maximum frequency of
harmonic vibrations is determined by the \mbox{relation
\cite{13}}\looseness=1
%5
\begin{equation}
\omega _{\rm ph}=\frac{2\pi }{d}\left(\! {\frac{E_{m}}{\rho
}}\!\right) ^{\!1/2}=\frac{2}{d}\, v, \label{eq5}
\end{equation}%

\noindent where $E_{m}$ is Young's modulus, $v$ is the thermal
velocity, $\rho =m/V_{i}$ is the substance density, $m$ is the
atomic (molecular) mass, and $V_{i}$ is the atomic (molecular)
volume. In vicinities of defect centers, $E_{m}$ becomes several
times smaller, and $\rho $ increases as the ratio between the
densities of defect components. The lattice constant $d$ increases
as the ratio between the atomic sizes $r$ of film components,
$r($GaAs)$/r($InGaAs). Hence, according to the estimations made for
the InGaAs heterostructure, $(d_{V}/d_{D})\sim 1/2$ and $\rho
_{v}/\rho _{D}\sim m_{V}/m_{D}\sim (30/45)$. Whence,
%6
\begin{equation}
\label{eq6} \frac{\omega _D }{\omega _V }=\left( \!{\frac{d_V }{d_D
}}\! \right) \left( \!{\frac{E_D }{E_V }\, \frac{S_V }{S_D }}
\!\right)^{\!1/2}\!=\frac{1}{2} \left(\! {\frac{1}{2}\cdot
\frac{1}{2}}\! \right)^{\!1/2}\!\sim \frac{1}{4}.
\end{equation}

Therefore, we may expect that the energy of local phonons in InGaAs
is several times lower than that in the bulk, i.e.
$E_{\mathrm{ph}}\sim \frac{1}{4}E_{\mathrm{ph}V}\sim 8\div
10~\mathrm{ meV}$.

When comparing the values obtained for $E_{\mathrm{ph}}$,
$E_{\mathrm{ex}}$, and $N$ with the growth parameters of specimens,
we may assert that the specimens with small values of $N$ have the
highest quality, i.e. the specimens with the highest mobility and
the one-mode phonon \mbox{spectrum.}\looseness=1\vspace*{2mm}

\section{Conclusions}

Heterostructures GaAs/In$_{x}$Ga$_{1-x}$As/GaAs that have a single quantum well
and are characterized by various growth parameters were studied with the use of the method
of low-temperature photoluminescence. The following facts were revealed.

(i) The photoluminescence spectra of examined specimens demonstrate
intense radiation bands. These bands are induced by the
recombination of excitons in the quantum well in a temperature
interval of 5--40~$ \mathrm{K}$ and by the recombination of free
charge carriers in the quantum well in the interval
$T=50\div200~\mathrm{K}$.

 (ii) For all specimens, the temperature dependence of the
photoluminescence intensity maximum position has an S-like shape. In
the low-temperature interval, the values calculated within the
Varshni model considerably deviate from the experimen- \mbox{tal
data.}\looseness=1

 (iii) The researched specimens revealed both monotonous and
nonmonotonous dependences of the photoluminescence band half-width
on the \mbox{temperature.}

 The features observed in the temperature dependences of
the maximum position and the half-width of the PL band testify that
all examined specimens contain defect states, in one quantity or
another, induced by fluctuations in the QW composition; in
particular, the inhomogeneities may occur owing to the segregation
of In-phase clusters (in the form of 3D islands). The values of
$E_{\rm ph}$ determined from the temperature dependences of the PL
band width turned out several times lower than the corresponding
characteristic values for bulk and surface phonons for all studied
specimens. Specimens with a high intensity of radiation emission and
a narrow radiation band were found to be characterized by a small
value of the Huang--Rhys factor and a one-mode phonon spectrum.
Local phonons of two types -- with energies of 3.5--4 and
18--19~meV, respectively -- take part in the process of exciton
scattering in specimens with low intensities of radiation and wide
radiation bands (i.e. with a worse structural
\mbox{quality}).\looseness=1

\vskip3mm

{\it The authors express their sincere gratitude to Corresponding
Member of the NAS of Ukraine V.G.~Litovchenko for the discussion of
the results of this work and useful advices.} %\vspace*{2mm}

%\vspace*{-3mm}
\rezume{%
Н.М. Литовченко, Д.В. Корбутяк, О.М. Стрільчук}{ЕКСИТОННІ
ХАРАКТЕРИСТИКИ In$_{x}$Ga$_{1-x}$As--GaAs\\ ГЕТЕРОСТРУКТУР З
КВАНТОВИМИ ЯМАМИ\\ ПРИ НИЗЬКИХ ТЕМПЕРАТУРАХ} {Проведена оцінка
характеристик гетероструктур з одиночною квантовою ямою
GaAs/In$_{x}$Ga$_{1-x}$As/GaAs з різними ростовими параметрами за
результатами вимірювань низькотемпературних спектрів
фотолюмінесценції (ФЛ), з відповідним теоретичним аналізом.
Проаналізовані експериментально отримані температурні залежності
енергії максимуму смуги ФЛ ($h\nu _{\max})$, півширини ($W_{0})$ та
інтенсивності $I$. Визначено параметри $E_{\mathrm{ph}}$ (енергія
локальних фононів), $E_{\mathrm{ex}}$ (енергія зв'язку екситонів) та
$N$ (фактор Хуанга--Ріс). Проведене зіставлення отриманих значень
$E_{\mathrm{ph}}$, $E_{\mathrm{ex}}$ та $N$ з ростовими параметрами
зразків дає підставу стверджувати, що найбільш якісними є %\linebreak
зразки з малим значенням $N$ і одномодовим фононним
\mbox{спектром.}}


\begin{thebibliography}{99}
\bibitem{1} I.A. Avrutskii, V.A. Sychugov, and B.A. Usievich, Fiz. Tekh.
Poluprovodn. \textbf{25}, 1787 (1991).\vspace*{0.5mm}

\bibitem{2} I.A. Avrutskii and V.G. Litovchenko, Fiz. Tekh. Poluprovodn.
\textbf{31}, 875 (1997).\vspace*{0.5mm}

\bibitem{3} M.M.\,Grigoriev,\,E.G.\,Gule,\,A.I.\,Klimovska,\,Yu.A.\,Ko\-rus, and
V.G. Litovchenko, Ukr. Fiz. Zh. \textbf{45}, 853
(2000).\vspace*{0.5mm}

\bibitem{4} I.A. Avrutskii, O.P. Osaulenko, V.G. Plotnichenko, and
Yu.N. Pyrkov, Fiz. Tekh. Poluprovodn. \textbf{26}, 1907
(1992).\vspace*{0.5mm}

\bibitem{5} H.D. Sun, R. Macaluso, S. Calvez, and M.D. Dawson, J. Appl.
Phys. \textbf{94}, 7581 (2003).\vspace*{0.5mm}

\bibitem{6} N.V. Kryzhanovskaya, A.Yu. Egorov, V.V. Mamutin, N.K. Po\-lyakov,
A.F. Tsatsulnikov, A.R. Kovsh, N.N.~Le\-dentsov, V.M. Ustinov, and
D. Bimberg, Fiz. Tekh. Poluprovodn. \textbf{39}, 735
(2005).\vspace*{0.5mm}

\bibitem{7} F.-I. Lai, S.Y. Kuo, J.S. Wang, R.S. Hsiao, H.C.~Kuo, J. Chi,
S.C. Wang, H.S. Wang, C.T. Liang, and Y.F.~Chen, J. Cryst. Growth
\textbf{291}, 27 (2006).\vspace*{0.5mm}

\bibitem{8} M. Soltani, M. Certier, R. Evrard, and E. Kartheusev, J. Appl.
Phys. \textbf{78}, 5626 (1995).\vspace*{0.5mm}

\bibitem{9} S.I. Pekar, Zh. Eksp. Teor. Fiz. \textbf{20}, 510 (1950).\vspace*{0.5mm}

\bibitem{10} C.J. Hwang, Phys. Rev. \textbf{180}, 827 (1969).\vspace*{0.5mm}

\bibitem{11} V.G. Litovchenko, N.L. Dmitruk, D.V. Korbutyak, and
A.V. Sarikov, Fiz. Tekh. Poluprovodn. \textbf{36}, 447
(2002).\vspace*{0.5mm}

\bibitem{12} V.A. Zuev, D.V. Korbutyak, V.G. Litovchenko, and A.V. Drazhan,
Fiz. Tverd. Tela \textbf{17}, 3300 (1975).\vspace*{0.5mm}

\bibitem{13} I. Bolesta, \textit{Solid State Physics} (Lviv, Lviv. Nats.
Univ., 2003) (in Ukrainian).\vspace*{6mm}

\begin{flushright}
{\footnotesize Received 12.12.12.\\ Translated from Ukrainian by
O.I. Voitenko}
\end{flushright}
\end{thebibliography}
\end{document}